\documentclass[journal]{IEEEtran}
\usepackage{amssymb}
\usepackage{amsmath}
\usepackage{algorithm}
\usepackage{algorithmic}
\usepackage{color}
\usepackage{bm}
\usepackage{fancyhdr}

\usepackage{cite}      

\usepackage{graphicx}  

\usepackage{subfigure} 
\begin{document}


\title{System Design and Analysis for Energy-Efficient Passive UAV Radar Imaging System using Illuminators of Opportunity}


\author{Zhichao~Sun,~\IEEEmembership{Member,~IEEE,}
        Junjie~Wu,~\IEEEmembership{Member,~IEEE,}
        Gary G. Yen,~\IEEEmembership{Fellow,~IEEE,}
        Hang Ren,~\IEEEmembership{Student Member,~IEEE,}
        Hongyang An,~\IEEEmembership{Student Member,~IEEE,}
        Jianyu Yang,~\IEEEmembership{Member,~IEEE}

\thanks{Z. Sun, J. Wu, H. Ren, H. An and Jianyu Yang are with School of Information and Communication Engineering, University of Electronic Science and Technology of China, Chengdu, People's Republic of China, 611731 (e-mail:threadkite@sina.com).

G. G. Yen is with the School of Electrical and Computer Engineering, Oklahoma State University, Stillwater, MN 74078 USA (e-mail:gyen@okstate.edu).

This work was supported in part by the National Natural Science Foundation of China under Grant 61901088, Grant 61922023, Grant 61771113, and Grant 61801099, in part by the Postdoctoral Innovation Talent Support Program under Grant BX20180059, and in part by the China Postdoctoral Science Foundation under Grant 2019M65338 and Grant 2019TQ0052.}}

\maketitle

\thispagestyle{fancy}          
\fancyhead{}                      
\lhead{This work has been submitted to the IEEE for possible publication. Copyright may be transferred without notice, after which this version may no longer be accessible.}           
\chead{}
\rhead{}
\lfoot{}
\cfoot{\thepage}   
\rfoot{}
\renewcommand{\headrulewidth}{0pt}       
\renewcommand{\footrulewidth}{0.7pt}
\vspace{2cm}
\fancyfoot{}
\renewcommand{\footrulewidth}{0pt}       

\begin{abstract}
Unmanned aerial vehicle (UAV) can provide superior flexibility and cost-efficiency for modern radar imaging systems, which is an ideal platform for advanced remote sensing applications using synthetic aperture radar (SAR) technology. In this paper, an energy-efficient passive UAV radar imaging system using illuminators of opportunity is first proposed and investigated. Equipped with a SAR receiver, the UAV platform passively reuses the backscattered signal of the target scene from an external illuminator, such as SAR satellite, GNSS or ground-based stationary commercial illuminators, and achieves bi-static SAR imaging and data communication. The system can provide instant accessibility to the radar image of the interested targets with enhanced platform concealment, which is an essential tool for stealth observation and scene monitoring. The mission concept and system block diagram are first presented with justifications on the advantages of the system. Then, a set of mission performance evaluators is established to quantitatively assess the capability of the system in a comprehensive manner, including UAV navigation, passive SAR imaging and communication. Finally, the validity of the proposed performance evaluators are verified by numerical simulations.
\end{abstract}


\section{Introduction}
Synthetic aperture radar (SAR) is capable of high-resolution imaging in all-day and all-weather environments, which can be mounted on different platforms in order to satisfy distinct remote sensing and mission requirements, including spaceborne, airborne and missile-borne platforms.

Compared with other platforms, unmanned aerial vehicles (UAVs) provide better mission safety and maneuverability with less cost. The SAR systems mounted on UAV platforms have been intensively studied in the current literature, including system design\cite{4768730}, imaging methods\cite{6072230} and motion compensation\cite{6144725}. Instead of using its own illuminator, a SAR radar can use an external illuminator and form a passive SAR. Consequently, passive SAR allows for silent operations, which can in turn enhance the system survivability in military scenarios. Moreover, due to the lack of transmitter, the system cost, weight and energy consumption can be greatly reduced compared with the active SAR counterpart, which is suitable for light-weight UAV platforms. In \cite{8646795}, a passive SAR system with airborne receiver using DVB-T illumination was investigated. Signal processing techniques were discussed and experiments were conducted to validate the system feasibility.

This paper proposes an energy-efficient passive UAV SAR system, which combines the UAV SAR and passive SAR technology. The system and mission concept are first put forward with discussions on the system components. Then, the mission performance of the passive UAV SAR system is analyzed in detail. A set of mission performance is established to quantitatively assess the performance of the system. The proposed evaluator set offers the theoretical foundation for optimizing the mission performance of the energy-efficient passive UAV radar imaging system by dynamically adapting the adopted UAV path in 3-D terrain environments, which will be our future work.

\section{Proposed Energy-Efficient Passive UAV SAR System}
The proposed SAR system utilizes external illuminators, which can be chosen from satellite-borne SAR (LEO-SAR and future GEO/MEO-SAR\cite{Hobbs2014System}), global navigation satellite system (GNSS) or ground-based stationary commercial illuminators (e.g., FM radio and DVB-T).
\begin{figure}[htb]
\centering
\centering\includegraphics[width=0.35\textwidth]{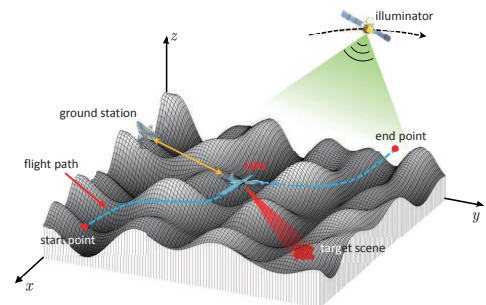}
\caption{A typical scenario for energy-efficient passive UAV SAR system.}\label{MissionConcept}
\end{figure}

The typical working scenario of the system in 3-D rough terrain is illustrated in Fig. \ref{MissionConcept}. Firstly, an optimal flight path is generated based on the digital elevation model (DEM) and motion parameters of the illuminator, which can safely guide the UAV to travel through the 3-D terrain and achieve passive bistatic SAR imaging and communication during the flight. The path is then loaded in the navigation system and the UAV is deployed according to the time schedule and flight path. The UAV platform travels from the start point to the end point and collects echo from the target scene during a predefined time window. Finally, after on-board signal preprocessing, the echo data is transmitted to the ground-based processing station for further imaging processing.

\begin{figure}[htb]
\centering
\centering\includegraphics[width=0.35\textwidth]{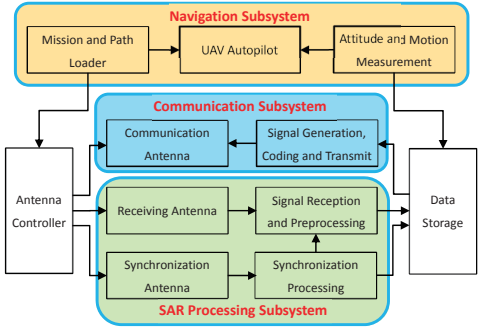}
\caption{Block diagram of the proposed system.}\label{blockdiagram}
\end{figure}

Based on the main functions described above, the block diagram of the system is shown in Fig. \ref{blockdiagram}. The system mainly consists of three functional modules, \emph{i.e.} navigation, communication and SAR processing subsystems, which collaborates with each other to accomplish the radar imaging task.

The navigation subsystem generates pilot commands and install in the mission and path loader for UAV autopilot. The SAR processing subsystem receives direct synchronization signal and echo signal reflected from the target scene. After synchronization and signal preprocessing, the data are stored in data storage for communication subsystem. An additional pair of antenna and corresponding receiving channel are equipped on the UAV for direct signal synchronization. Finally, the communication subsystem is dedicated to transmit the echo data, synchronization information and motion parameters to the ground processing station, where the following SAR imaging is finished. On-board SAR imaging processing is an energy and time consuming task. Therefore, for a miniaturized and light-weighted UAV platform, SAR imaging task is achieved at the ground processing station.

\section{Mission Performance Evaluator}

\textbf{(1) UAV Navigation and Energy Efficiency}

In this paper, energy consumption of UAV path in 3-D motion is derived to represent the flight performance instead of path length, which is a more practical and comprehensive evaluator for UAV navigation and flight performance. For the proposed system, the energy consumption of UAV flight control is the main part and is dependent on the adopted flight path. The UAV platform may require banked level turn, climbing and descending in a 3-D terrain. The continuous UAV path is modeled as a series of discrete path segments with $N_{dp}$ discrete points denoted as $({P_1}, \cdots ,{P_i}, \cdots ,{P_{{N_{dp}}}})$. Therefore, total energy consumption of an arbitrary UAV flight path $d(i)$ can be approximated by the sum of energy consumption during each discrete path segment.
\begin{align}
\label{EnergyConsumption}
E\left( {d(i)} \right) &= \sum\limits_{i = 1}^{{N_{dp}} - 1} {\left( {{P_{drag}}(i)T(i) + mg\left[ {h({P_{i + 1}}) - h({P_i})} \right]} \right)} \nonumber\\
 &+ \sum\limits_{i = 1}^{{N_{dp}} - 1} {\left( {\frac{1}{2}m\left[ {{{\left\| {v({P_{i + 1}})} \right\|}^2} - {{\left\| {v({P_i})} \right\|}^2}} \right]} \right)},
\end{align}
where ${h({P_i})}$ is the UAV height at $P_i$ and ${T(i)}$ is the time required for UAV to travel from $P_i$ to $P_{i+1}$. ${{P_{drag}}}$ is the power needed to overcome drag.
\begin{align}
\label{pdrag}
{P_{drag}} = {c_1}{\left\| {\vec v} \right\|^3} + \frac{{{c_2}}}{{\left\| {\vec v} \right\|}}\left( {\frac{{v_a^2}}{{{{\left\| {\vec v} \right\|}^2}}} + \frac{{{{\left\| {\vec a} \right\|}^2} - {{\left( {\frac{{{{\vec a}^T}\vec v}}{{\left\| {\vec v} \right\|}}} \right)}^2}}}{{{g^2}}}} \right).
\end{align}
The first term in (\ref{EnergyConsumption}) represents the energy consumption to overcome drag force, which is dependent on the velocity, centrifugal acceleration and climbing angle of the UAV during the flight. The second term is the gravitational potential energy, which is determined by the UAV mass and change of height during a specific time period. The third term is the kinetic energy related to the flight speeds of UAV at initial and final positions. Therefore, the energy consumption is dependent on the UAV path and a path with ascending trajectory generally consumes more energy.

In addition to the energy consumption performance, safety navigation of the flight path for passive UAV SAR should be considered in 3-D terrain scenario path planning. In this paper, we adopt the threat value $f_{threat}$ in \cite{7516590} to represent the safety performance of a flight path.

\textbf{(2) Passive UAV SAR Imaging Performance}

Spatial resolution and data size are considered to comprehensively evaluate the performance of the proposed system in practical applications. The spatial resolution changes with geometric location of the targets. The resolution cell area $S_c$ of the reference target can not fully represent the resolution performance of the entire target scene. Therefore, a modified resolution evaluator is proposed based on $S_c$, where the spatial variance of resolution is considered.
\begin{align}
\label{Modifiedresolution}
{\bar S_c}\left( {d(i)} \right) = \frac{{S_c^{\max }}}{{S_c^{\min }}} \cdot \frac{1}{{{N_s}}}\sum\limits_{j = 1}^{{N_s}} {S_c^j},
\end{align}
$N_s$ sampled points are evenly distributed in the ${W_r} \times {W_a}$ scene size and each point can represent the resolution performance of the targets in its vicinity without significant degradation. ${S_c^j}$ is the $S_c$ of the \emph{j}th sampled point. ${S_c^{\max }}$ and ${S_c^{\min }}$ are the maximum and minimum $S_c$ among the $N_s$ sampled points, respectively. In (\ref{Modifiedresolution}), the first part is disequilibrium factor, \emph{i.e. }the ratio of ${S_c^{\max }}$ to ${S_c^{\min }}$, which represents the uniformity of the distribution of resolution performance. Larger disequilibrium factor indicates greater spatial variance of resolution performance, which is not desired in practice.

On the other hand, the size of echo data matrix to be transmitted by the communication subsystem can be calculated as
\begin{align}
\label{AzimuthNum}
{N_a} &= \left( {{{W_a}/v + T_a}} \right) \cdot PRF, \nonumber\\
{N_r} &= \left( {\Delta {R_{bi}}/c} \right) \cdot {S_{rate}}
\end{align}
where ${N_a}$ and ${N_r}$ are the azimuth and range data sizes after range pulse compression, respectively. $PRF$ is the pulse repetition frequency and $T_a$ is the synthetic aperture time. ${\Delta {R_{bi}}}$ is the bi-static slant range variation within $W_r$ and $S_{rate}$ is range sampling rate. Therefore, the total size of the echo data matrix to be transmitted is ${D_{echo}}\left( {d(i)} \right) = 128{N_a} \cdot {N_r}$ in bits. The data size should be jointly considered with communication capacity to ensure complete transmission of echo data.

\textbf{(3) Communication Performance with Ground Station}

After echo collection and preprocessing, the data are transmitted to a ground-based processing station. Suppose the communication link between UAV platform and ground station is line-of-sight link and Doppler effects due to UAV motion has been compensated. Then, the communication channel can be expressed as a free-space path loss model. Therefore, the maximum amount of data bits that can be transmitted from the UAV to the ground station can be expressed as
\begin{align}
\label{databits}
{D_{com}}\left( {d(i)} \right) = \sum\limits_{i = i_{com}^{start}}^{i_{com}^{end}} {T(i) \cdot {B_{com}}{{\log }_2}\left( {1 + \frac{{{P_{com}}{\beta _0}}}{{{\sigma ^2}{l^2}(i)}}} \right)}.
\end{align}
where $i_{com}^{start}$ and $i_{com}^{end}$ denote the start and end indices of discrete path point in $d(i)$. $T(i)$ is the time required for UAV to travel through the \emph{i}-th discrete path. $B_{com}$ is the communication bandwidth. $\sigma$ is the receiver Gaussian white noise power and $P_{com}$ denotes the transmission power assumed to be constant during the path. The scaling factor ${\beta _0}{P_{com}}/{\sigma ^2}$ is the receiving signal to noise ratio at a reference distance, which is determined by the system parameters. The total amount of transmitted data ${D_{com}}\left( {d(i)} \right)$ during the flight should be larger than the data size of echo signal ${D_{echo}}\left( {d(i)} \right)$, to make sure all the data are transmitted to the ground station.

\section{Simulations}

The mission performance and their characteristics are analyzed and verified by numerical simulations in this section.
\begin{figure}[htb]
\centering
\centering\includegraphics[width=0.4\textwidth]{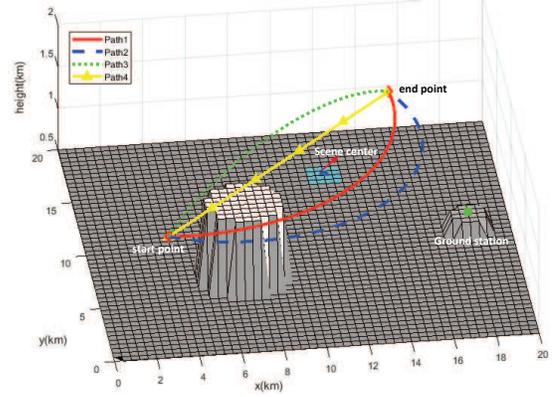}
\caption{Simulated scenario of passive UAV SAR system.}\label{simusinario}
\end{figure}

The simulated working scenario of the system is illustrated in Fig. \ref{simusinario}. The 3-D terrain contains 500m altitude 'flat ground' and two elevated areas, $(7,7,1.4)$km with 2km radius and $(18,10,0.7)$km with 1km radius. The interested target scene to be observed is located at $(12,16)$km on the 'flat ground', which is marked by blue squares. The ground processing station is located at $(18,10,0.72)$km. The coordinates of the start point and end point of the UAV paths are set as $(3,3.5,1.5)$km and $(15,15.6,1.5)$km, respectively. Four potential UAV paths are generated for performance evaluation. Path 1 and path 2 are horizontal arcs and path 3 is vertical arc. Path 4 is straight line. The system parameters are given in Table. \ref{systempara}, where a GEO-SAR is chosen as the illuminator.
\begin{table}[!htb]
\centering
\caption{Simulated parameters of passive UAV SAR system.}\label{systempara}
\centering\includegraphics[width=0.4\textwidth]{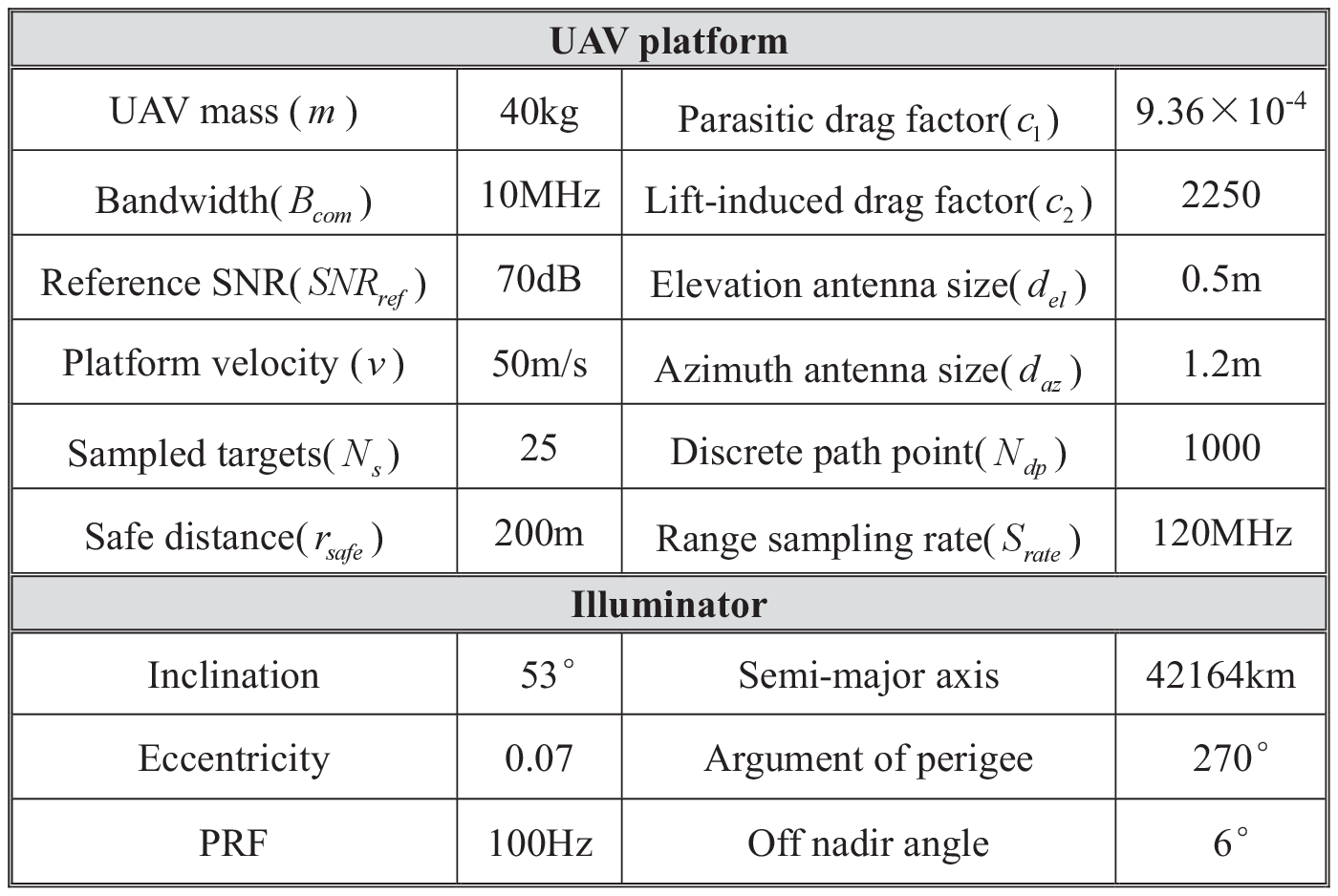}
\end{table}

\begin{table}[!htb]
\centering
\caption{Mission performance of the simulated paths.}\label{pathperform}
\centering\includegraphics[width=0.43\textwidth]{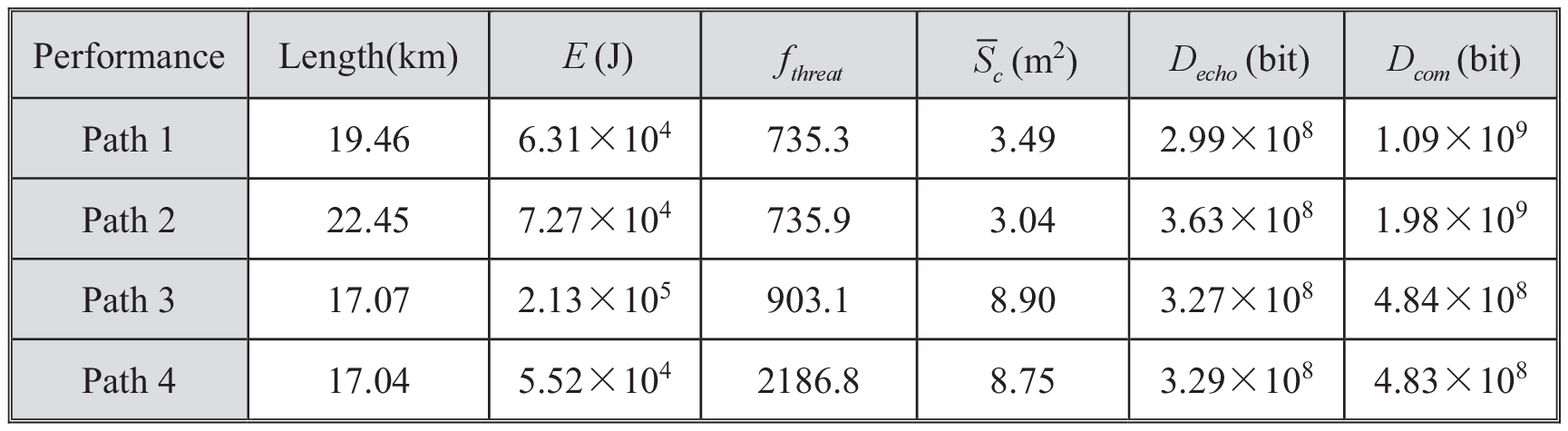}
\end{table}

The measured mission performance of the four paths in the simulated scenario is illustrated in Table \ref{pathperform}. The mission performance of the four UAV paths in the simulated scenario is analyzed and verified in three aspects in the following.

\textbf{UAV navigation:} It can be observed from Table \ref{pathperform} that the straight-line path, \emph{i.e.} path 4, presents the shortest path length with smallest energy consumption. Although the lengths of path 3 is only 0.03km longer than that of path 4, the energy consumption is significantly larger due to the climbing motion of UAV platform. Therefore,it can be concluded that a 'smoother' path with less maneuvers in elevation direction is desired for energy-efficient UAV SAR mission. On the other hand, the threat value of path 1 is the smallest while path 4 is the largest. The path should keep enough distance from the terrain obstacles to guarantee the safe navigation for UAV.

\begin{figure}[htp]
\begin{center}
\subfigure[Imaging result of path 1.]{\label{path1Imaging}\includegraphics[scale=0.36]{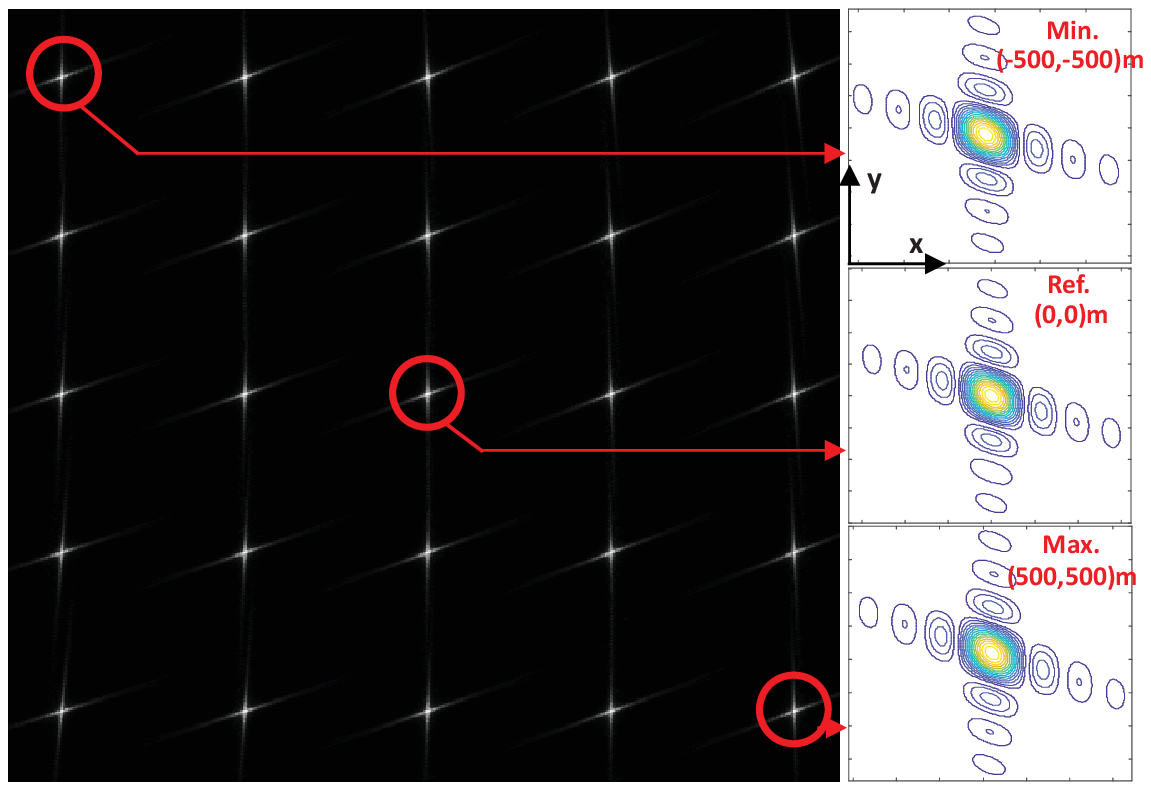}}
    \hfil
    \subfigure[Imaging result of path 2.]{\label{path2Imaging}\includegraphics[scale=0.36]{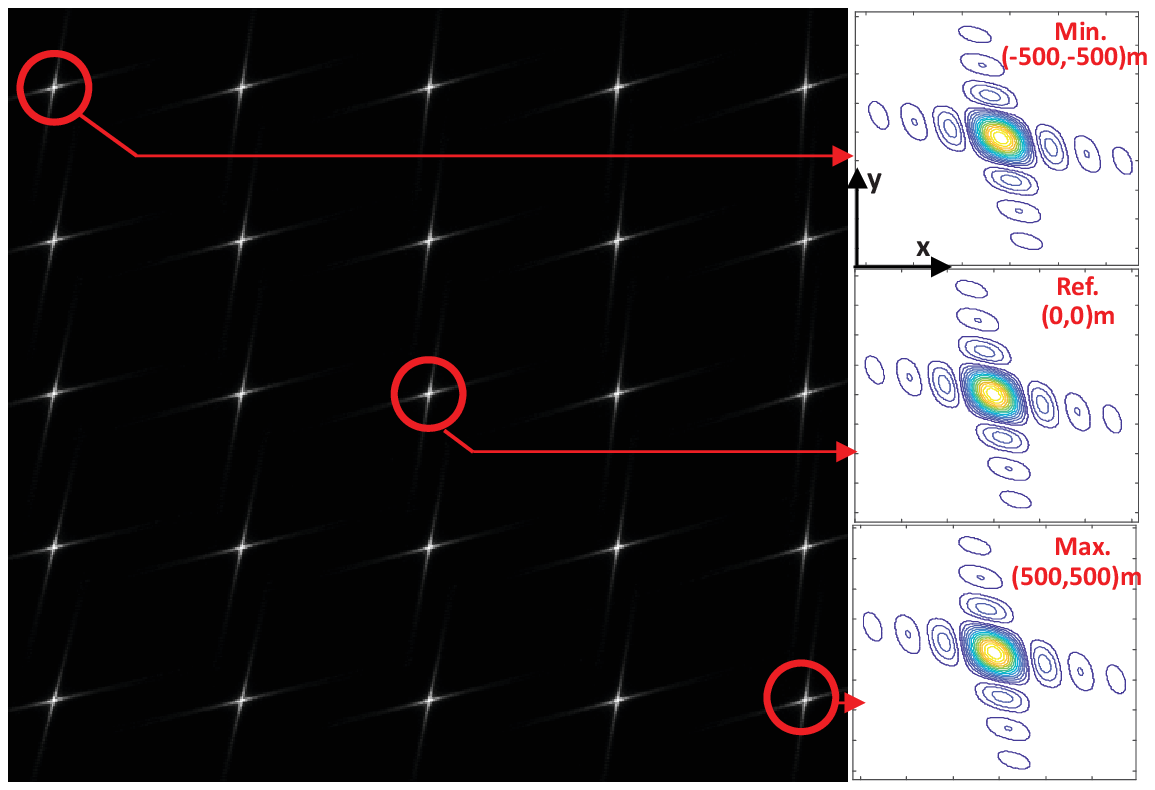}}
        \hfil
    \subfigure[Imaging result of path 3.]{\label{path3Imaging}\includegraphics[scale=0.35]{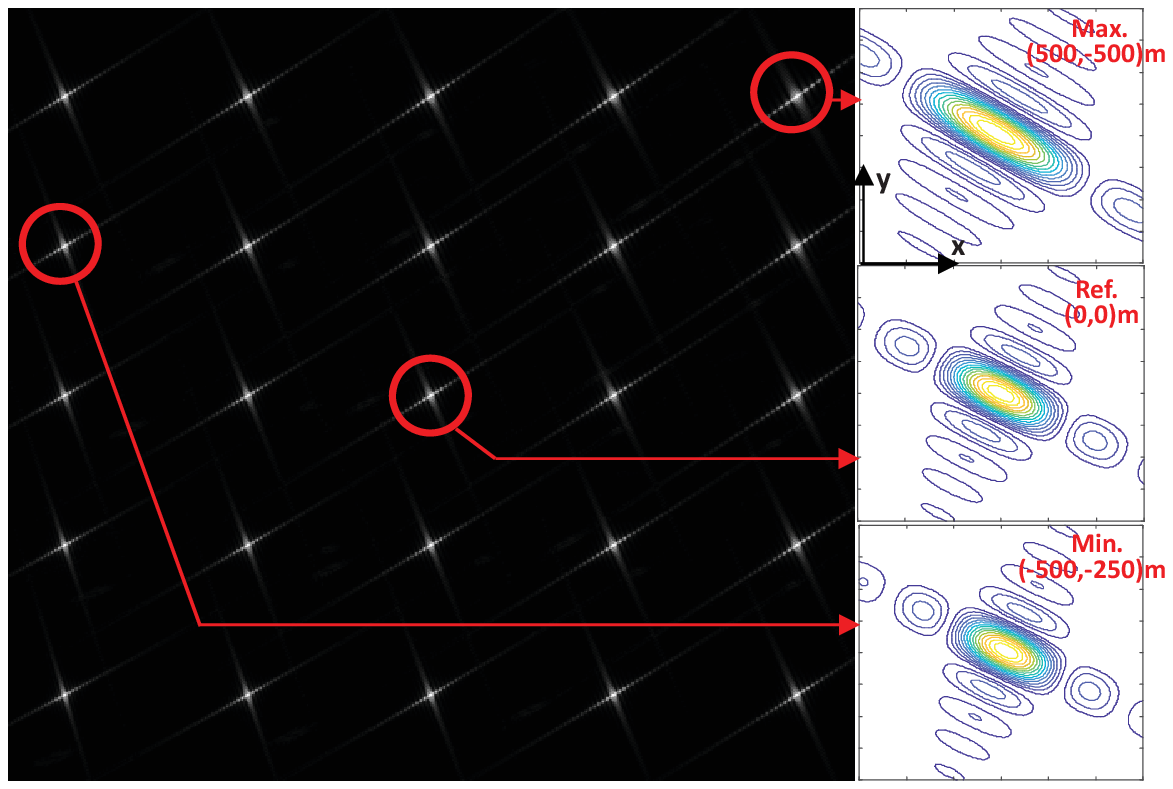}}
            \hfil
     \subfigure[Imaging result of path 4.]{\label{path4Imaging}\includegraphics[scale=0.36]{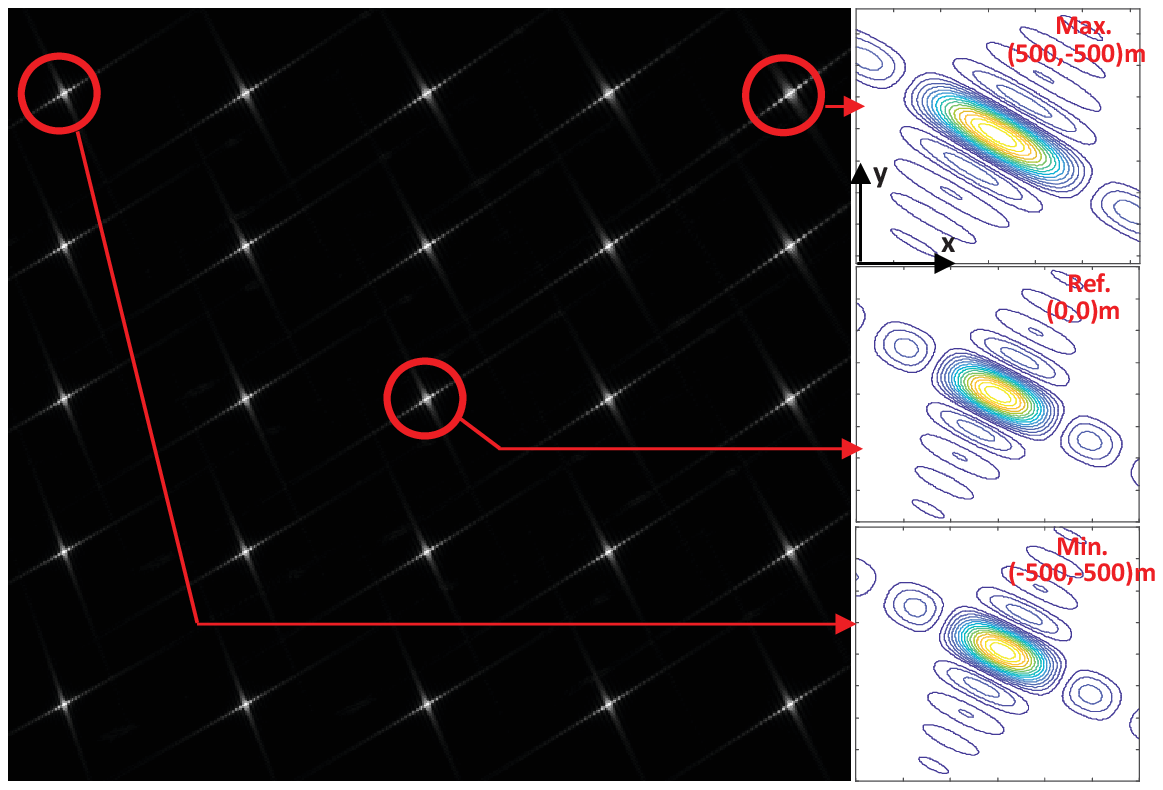}}
  \end{center}
  \caption{Imaging results.}
  \label{imagingresult}
\end{figure}

\textbf{Passive SAR imaging:}
From Table \ref{pathperform}, path 1 and path 2 can provide better imaging performance compared with path 3 and path 4. SAR raw-data simulation results are given in Fig. \ref{imagingresult}, where 25 ideal point targets are evenly distributed in the $1km \times 1km$ imaging scene. Three selected targets include the reference point '$Ref.$' and the targets with minimum and maximum $S_c$, denoted as '$Min.$' and '$Max.$', respectively. The coordinates of the selected targets are given and the contours of the imaging results are plotted within $16m \times 12m$ ground xy plane. The measured performance metrics are shown in Table \ref{imageIRW}, which validate the imaging performance evaluator ${\bar S_c}$ in Table \ref{pathperform}.

\begin{table}[!htb]
\centering
\caption{Measured imaging metrics of the selected targets.}\label{imageIRW}
\centering\includegraphics[width=0.35\textwidth]{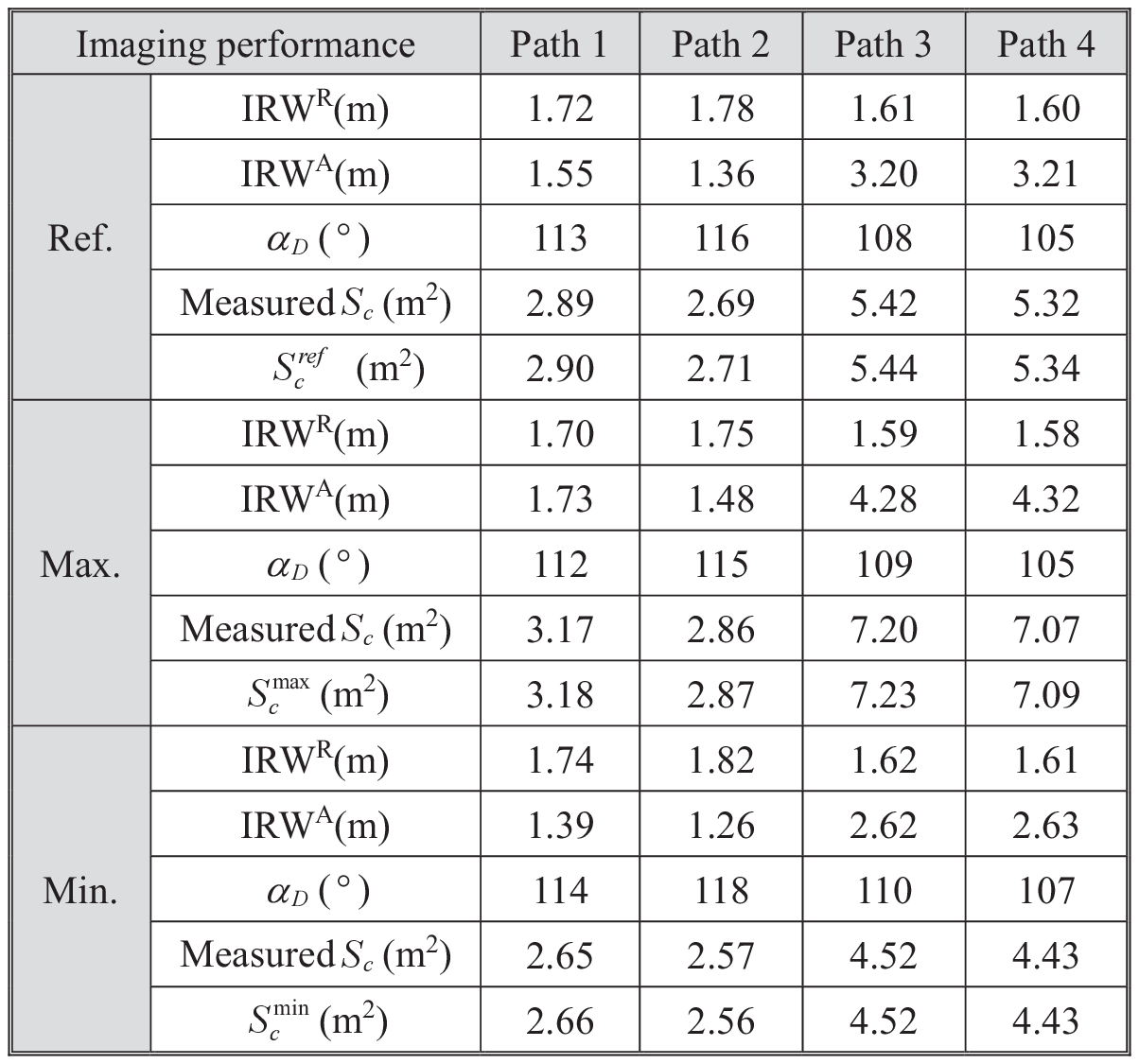}
\end{table}

\textbf{Communication:} Inspecting Table \ref{pathperform}, all of the four paths can completely transmit the echo data to the ground processing station. However, path 1 and path 2 can provide better communication performance with higher data transmission capacity compared with path 3 and path 4. It can be observed from Fig. \ref{simusinario} that the detours of path 1 and path 2 shorten the distance between the path and the ground station, which enhances the communication performance. The abundance of communication capacity enables more mission opportunities during the flight, such as additional observation task and data retransmission.

Based on the above analysis, path 1 can provide better overall mission performance, with satisfactory imaging performance, energy efficiency and safety. Path 2 can achieve better imaging and communication performance at the cost of longer path and more energy consumption. Both paths can be adopted to accomplish the simulated SAR mission.

\section{Conclusion}
In this paper, an energy-efficient passive UAV SAR system using external illuminator is put forward. The system concept and block diagram are firstly analyzed. Then, the mission performance evaluators are analyzed and verified by numerical simulations in a 3-D terrain environment, which can accurately represent the system performance and lays the theoretical foundation for mission planning.

\bibliographystyle{IEEEtran}
\bibliography{monoSQ}
\end{document}